\documentclass[twocolumn]{aastex631}

\usepackage{amsmath}

\usepackage{xcolor}

\newcommand\R[1]{\textcolor{orange}{#1}}

\newcommand{\correction}[1]{\textcolor{black}{#1}}

\usepackage{comment}

\usepackage{natbib}
\usepackage{upgreek}
\defcitealias{TJCI2024}{TJCI2024}

\begin{document}

\title{Stellar Contamination Correction Using Back-to-Back Transits of TRAPPIST-1~b and c}

\shorttitle{Back-to-back transits to correct for TLS}
\shortauthors{Rathcke et al.}

\correspondingauthor{Alexander D. Rathcke}
\email{rathcke@space.dtu.dk}

\author[0000-0002-4227-4953]{Alexander D. Rathcke}
\affil{Department of Space Research and Space Technology, Technical University of Denmark, Elektrovej 328, 2800 Kgs. Lyngby, Denmark}

\author[0000-0003-1605-5666]{Lars A. Buchhave}
\affil{Department of Space Research and Space Technology, Technical University of Denmark, Elektrovej 328, 2800 Kgs. Lyngby, Denmark}

\author[0000-0003-2415-2191]{Julien de Wit}
\affil{Department of Earth, Atmospheric and Planetary Sciences, Massachusetts Institute of Technology, 77 Massachusetts Avenue, Cambridge, MA 02139, USA}
\affil{Kavli Institute for Astrophysics and Space Research, Massachusetts Institute of Technology, Cambridge, MA 02139, USA}

\author[0000-0002-3627-1676]{Benjamin V. Rackham}
\affil{Department of Earth, Atmospheric and Planetary Sciences, Massachusetts Institute of Technology, 77 Massachusetts Avenue, Cambridge, MA 02139, USA}
\affil{Kavli Institute for Astrophysics and Space Research, Massachusetts Institute of Technology, Cambridge, MA 02139, USA}

\author[0000-0003-3829-8554]{Prune C. August}
\affil{Department of Space Research and Space Technology, Technical University of Denmark, Elektrovej 328, 2800 Kgs. Lyngby, Denmark}

\author[0000-0001-8274-6639]{Hannah Diamond-Lowe}
\affil{Department of Space Research and Space Technology, Technical University of Denmark, Elektrovej 328, 2800 Kgs. Lyngby, Denmark}

\author[0000-0002-6907-4476]{João M. Mendonça}
\affil{Department of Space Research and Space Technology, Technical University of Denmark, Elektrovej 328, 2800 Kgs. Lyngby, Denmark}
\affil{Department of Physics and Astronomy, University of Southampton, Highfield, Southampton SO17 1BJ, UK}
\affil{School of Ocean and Earth Science, University of Southampton, Southampton, SO14 3ZH, UK}

\author[0000-0003-3355-1223]{Aaron Bello-Arufe}
\affil{Jet Propulsion Laboratory, California Institute of Technology, Pasadena, CA 91109, USA}
\affil{Department of Space Research and Space Technology, Technical University of Denmark, Elektrovej 328, 2800 Kgs. Lyngby, Denmark}

\author[0000-0003-3204-8183]{Mercedes L\'opez-Morales}
\affil{Space Telescope Science Institute, 3700 San Martin Dr, Baltimore MD 21218, USA}

\author[0000-0003-4269-3311]{Daniel Kitzmann}
\affil{Space Research and Planetary Sciences, Physics Institute, University of Bern, Gesellschaftsstrasse 6, 3012 Bern, Switzerland}
\affil{Center for Space and Habitability, University of Bern, Gesellschaftsstrasse 6, 3012 Bern, Switzerla}

\author[0000-0003-1907-5910]{Kevin Heng}
\affil{Ludwig Maximilian University, Faculty of Physics, University Observatory, Scheinerstr. 1, Munich D-81679, Germany}
\affil{ARTORG Center for Biomedical Engineering Research, University of Bern, Murtenstrasse 50, CH-3008, Bern, Switzerland}
\affil{University College London, Department of Physics \& Astronomy, Gower St, London, WC1E 6BT, United Kingdom}
\affil{University of Warwick, Department of Physics, Astronomy \& Astrophysics Group, Coventry CV4 7AL, United Kingdom}

\begin{abstract}

\correction{Stellar surface heterogeneities, such as spots and faculae, often contaminate exoplanet transit spectra, hindering precise atmospheric characterization. We demonstrate a novel, epoch-based, model-independent method to mitigate stellar contamination, applicable to multi-planet systems with at least one airless planet. We apply this method using quasi-simultaneous transits of TRAPPIST-1~b and TRAPPIST-1~c observed on July 9, 2024, with JWST/NIRSpec PRISM. These two planets, with nearly identical radii and impact parameters, are likely either bare rocks or possess thin, low-pressure atmospheres, making them ideal candidates for this technique, as variations in their transit spectra would be primarily attributed to stellar activity. Our observations reveal their transit spectra exhibit consistent features, indicating similar levels of stellar contamination. We use TRAPPIST-1~b to correct the transit spectrum of TRAPPIST-1~c, achieving a 2.5$\times$ reduction in stellar contamination at shorter wavelengths. At longer wavelengths, lower SNR prevents clear detection of contamination or full assessment of mitigation. Still, out-of-transit analysis reveals variations across the spectrum, suggesting contamination extends into the longer wavelengths. Based on the success of the correction at shorter wavelengths, we argue that contamination is also reduced at longer wavelengths to a similar extent. This shifts the challenge of detecting atmospheric features to a predominantly white noise issue, which can be addressed by stacking observations. This method enables epoch-specific stellar contamination corrections, allowing co-addition of planetary spectra for reliable searches of secondary atmospheres with signals of 60–250~ppm. Additionally, we identify small-scale cold ($\sim$2000 K) and warm ($\sim$2600 K) regions almost uniformly distributed on TRAPPIST-1, with overall covering fractions varying by $\sim$0.1\% per hour.}

\end{abstract}

\keywords{Transmission spectroscopy (2133); Stellar atmospheres (1584); Planet hosting stars (1242); Exoplanet atmospheres (487); Fundamental parameters of stars (555); Starspots (1572)}

\section{Introduction} \label{sec:intro}

JWST is revolutionizing the study of exoplanets and their atmospheres, providing unprecedented insights into worlds beyond our solar system \citep[e.g.,][]{JTECERST2023, Ahrer2023, Alderson2023, Feinstein2023, Rustamkulov2023_ERS}. However, to fully harness the capabilities of JWST, the astronomical community must overcome the significant challenge of stellar contamination \citep[e.g.,][]{sag21,lim:2023,moran:2023,rackham:2024}. This challenge arises because transmission spectra, which probe exoplanetary atmospheres, are shaped by both the planet and its host star, with the stellar brightness distribution contributing to the observed signal \citep[e.g.,][]{Rackham2017, Rackham2018, Rackham2019}. If these stellar and planetary signals are not accurately disentangled, the resulting atmospheric analysis of the planet can be significantly biased \citep{iyer:2020}. Despite its importance, separating the stellar contribution from the planetary signal in transmission spectra remains an unresolved and complex challenge.

\begin{figure*}[ht!]
	\centering
	\includegraphics[width=0.95\textwidth]{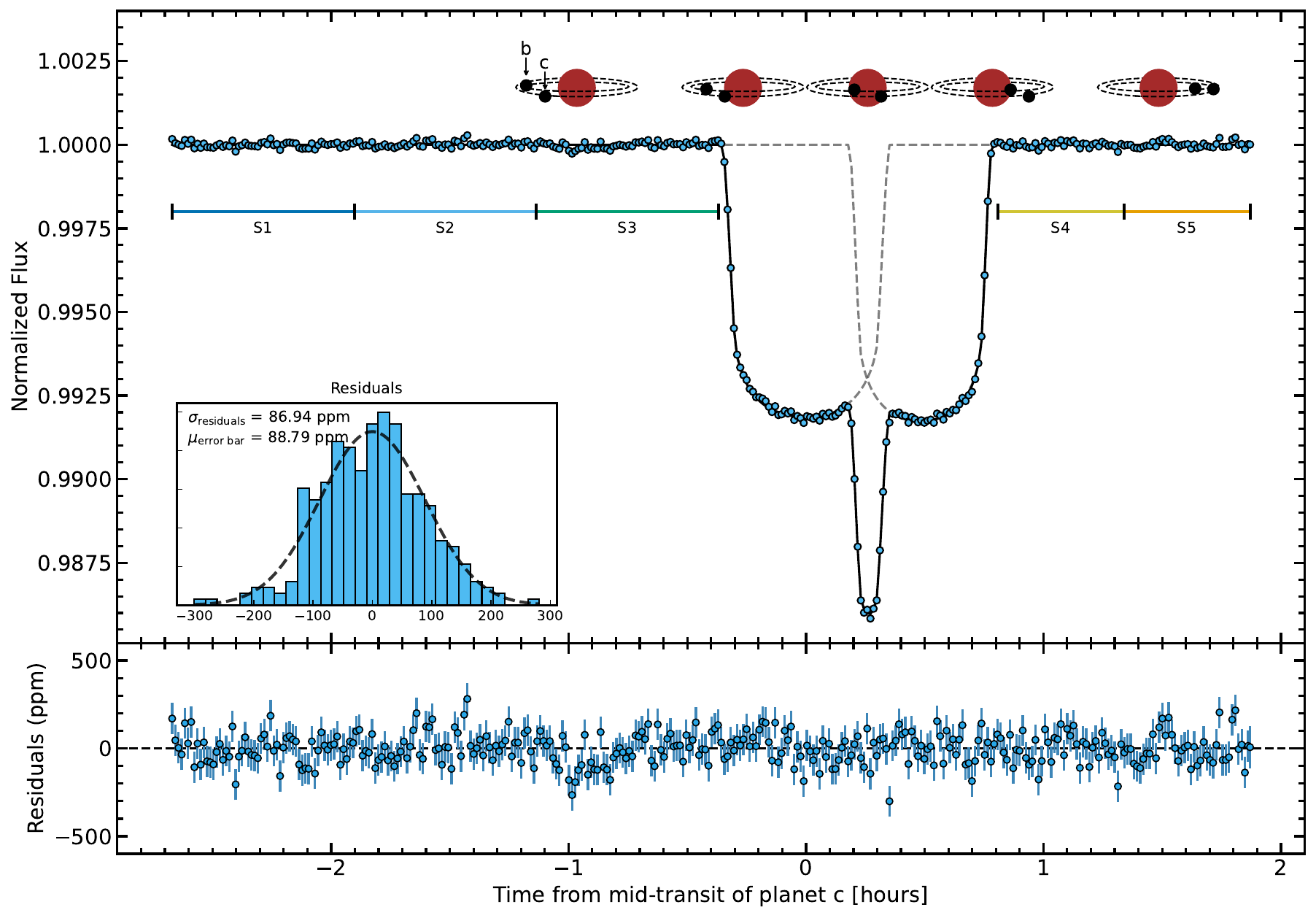}
	\caption{Normalized white light curves of TRAPPIST-1~c and TRAPPIST-1~b observed with JWST/NIRSpec PRISM during their quasi-simultaneous transits on July 9, 2024. \textbf{Top panel:} A schematic of the geometry of the quasi-simultaneous transits is displayed at the top. The blue circles represent the data points after removing the systematics model (a combination of a linear slope and a Gaussian Process, see \autoref{sec:in-transit_spec}). The solid black line indicates the fitted transit model. The data have been binned in time from the original 10,208 integrations to the 341 points shown here to accommodate the computation of the Gaussian Process systematics model. Horizontal lines labeled S1 through S5 mark the time bins used in the out-of-transit stellar analysis (see \autoref{sec:oot_spectra_analysis}). The inset displays a histogram of the residuals after subtracting the combined transit and systematics model, compared to a Gaussian distribution with a standard deviation equal to the mean measurement error plus a fitted white noise term. \textbf{Bottom panel:} Residuals after subtracting both the transit and systematic models.}
	\label{fig:white_light}
\end{figure*}

The primary obstacle to addressing the challenge of stellar contamination is the limited fidelity of current stellar models \citep{wakeford:2019,garcia:2022,rackham:2024}. This lack of fidelity can significantly hinder the reliable determination of the number of components on the stellar surface (e.g., spots and faculae). As a result, it leads to a wide range of possible corrections, with associated uncertainties that can be much larger than the photon noise \citep{wakeford:2019,garcia:2022}. Fortunately, models with sufficient fidelity can allow for an accurate determination of the true number of stellar components \citep{rackham:2024}. Both theory-driven \citep{Witzke2021,rackham:2024} and data-driven efforts \citep[][hereafter \citetalias{TJCI2024}]{Berardo2024,TJCI2024} are now focused on developing the next generation of stellar models with the fidelity required for the JWST era.

As the community advances toward the development of a new generation of stellar models, an alternative approach has been suggested: using the quasi-simultaneous transit of an airless planet to capture the stellar signal, thereby providing an epoch-specific, model-independent correction for stellar contamination \citepalias{TJCI2024}. However, this technique requires further testing to establish best practices for its reliable application. To address this, on July 9th, 2024, we observed the quasi-simultaneous transits of TRAPPIST-1~b and~c \citep{Gillon2016,Gillon2017} using JWST NIRSpec PRISM. 

TRAPPIST-1~b and~c have a similar size and impact parameter \citep{Agol2021}, and both are believed to have, at most, thin, low-surface-pressure atmospheres \citep{Greene2023, Zieba2023, GillonInReview}. Consequently, their transit spectra are expected to primarily show stellar contamination and, if obtained quasi-simultaneously and if the two planets traverse the same transit cord across the star, should be indistinguishable. If the transit spectra of these planets display discrepancies smaller than the atmospheric signals expected for secondary atmospheres \correction{(ranging from $\sim$60 ppm for the temperate planets to $\sim$250 ppm for the hotter planets, e.g., \citealt{Lincowski2018}, \citealt{Fauchez2019},\citealt{lim:2023}, \citealt{triaud2023}, \citetalias{TJCI2024})}, it would suggest that correcting for stellar contamination using back-to-back transits can sufficiently mitigate stellar activity allowing reliable searches for secondary atmospheres around terrestrial exoplanets in multi-planet systems with at least one planet without an atmosphere. Moreover, this would imply that both TRAPPIST-1~b and~c can serve as anchor planets in the continued search for an atmosphere within the TRAPPIST-1 system, thereby increasing the number of opportunities for multitransit observations. 

Here, we apply this technique to TRAPPIST-1~b and TRAPPIST-1~c, utilizing their quasi-simultaneous transits observed with JWST to assess whether their stellar contamination signals are similar enough to enable a robust correction for stellar activity.

We present data acquisition and reduction in \correction{Section \ref{sec:data}, followed by the methods and their respective results in Section \ref{sec:meth_and_res}. Our findings are discussed in Section \ref{sec:discussion} and conclusions are drawn in Section \ref{sec:conclusion}.}

\section{Data Acquisition and Reduction} \label{sec:data}

The final visit of the JWST Cycle 1 program GO 2420 (PI: Rathcke), which aims to evaluate the presence of an atmosphere around TRAPPIST-1~c, was strategically scheduled for July 9, 2024\footnote{The JWST data used in this paper can be found in MAST: \dataset[10.17909/4zcj-k904]{http://dx.doi.org/10.17909/4zcj-k904}}, to coincide with the quasi-simultaneous transit of TRAPPIST-1~b following \citetalias{TJCI2024}. Observations were carried out using the NIRSpec PRISM instrument, utilizing the S1600A1 slit, the 32-pixel wide SUB512 subarray and the NRSRAPID readout pattern. To optimize the signal-to-noise ratio, particularly at the red end of the spectrum, we employed six groups per integration. This configuration resulted in partial saturation of a small number of pixels near the peak of the stellar SED.

Data reduction was performed using our custom-built pipeline, \texttt{Frida}. \texttt{Frida} is a completely independently developed JWST pipeline only utilizing parts of stage 1 from \texttt{jwst}\footnote{\url{https://github.com/spacetelescope/jwst}}, the official STScI JWST pipeline. The pipeline follows the stage 1 TSO steps and additionally masks out entire columns if any pixel saturates at a given group, and removes any subsequent groups. Additionally, \texttt{Frida} includes routines to remove pre-amplifier reset noise at the group level and to eliminate column-wise 1/f noise. We employed a custom pixel mask to identify and interpolate bad pixels. Instead of performing up-the-ramp fitting, we calculate the rate by subtracting the first group from the last group of each ramp. We find this approach to be more robust than ramp fitting when the data contains only a few groups, as using the ramp fitting routine from \texttt{jwst} produced similar but slightly noisier results.

Starting from stage 2, \texttt{Frida} does not utilize any routines from \texttt{jwst}. We utilize a custom bad pixel map masking out manually identified bad pixels in addition to the ones identified in the \correction{data quality} map from stage 1. Cosmic rays are identified and removed using time-series analysis to detect 5-sigma outliers at the pixel level light curves, which were then replaced with values smoothed by a Gaussian filter with a length of 10 integrations. Finally, we perform optimal extraction of the spectra by using a normalized smoothed median-weighted spectrum (representing the PSF) to define pixel weights. The resulting white-light curve, obtained by summing the extracted spectrum over the wavelength range from 0.87 $\mu\text{m}$ to
 4.37 $\mu\text{m}$, is presented in \autoref{fig:white_light}.

\section{Methods and Results} \label{sec:meth_and_res}

Various methodologies were employed to analyze both the in-transit and out-of-transit data to characterize stellar activity signals. We first present the method used for the out-of-transit data and the insights gained, then turn to the in-transit data.

\subsection{Out-of-transit spectra}\label{sec:oot_spectra_analysis}

\begin{figure*}[htp]
    \centering
    \includegraphics[width=\textwidth]{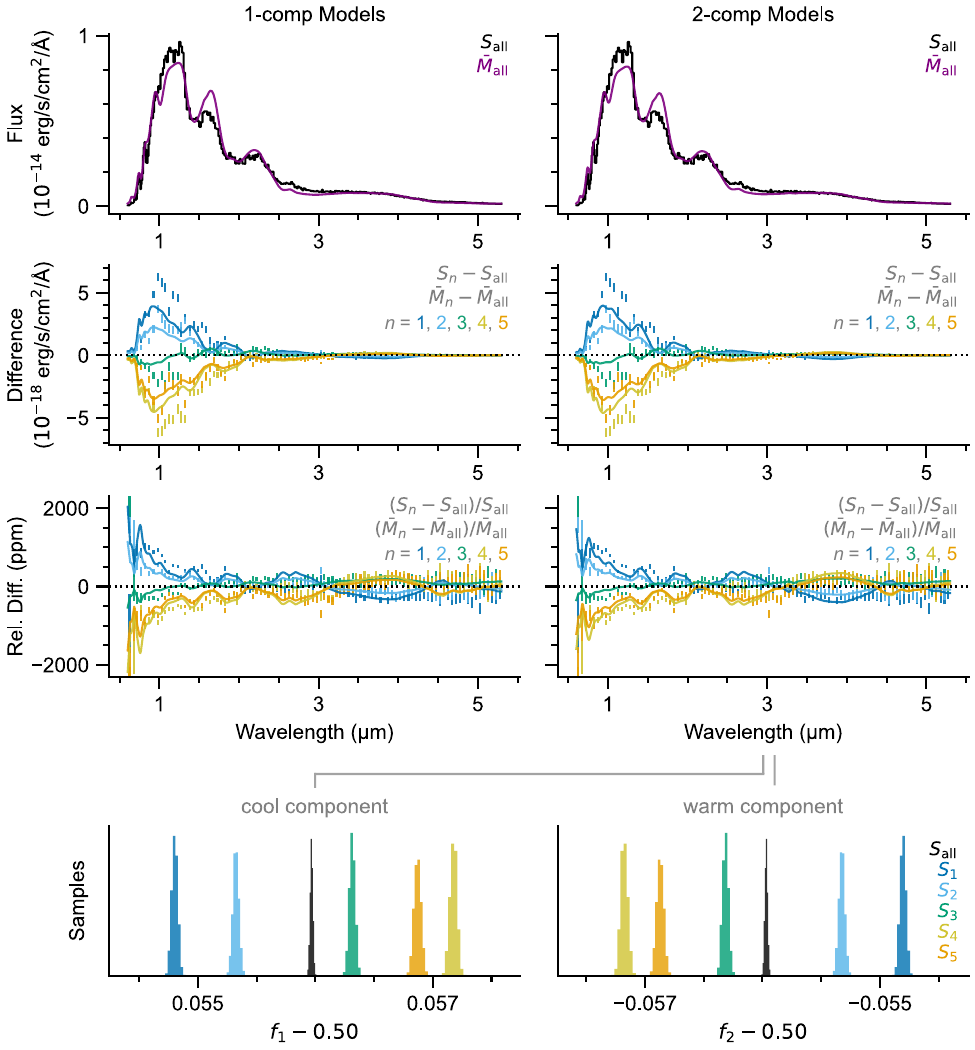}
    \caption{
        \correction{Out-of-transit stellar spectra and model fits. 
\textbf{Top row:} Mean stellar spectrum for all out-of-transit data ($S_\mathrm{all}$, which combines the $S_1$ through $S_5$ stellar spectra, as defined by the out-of-transit segments shown in Fig.~\ref{fig:white_light}) alongside the 1-component (left) and 2-component (right) model fits ($\bar{M}_\mathrm{all}$). Both fits are qualitatively similar, as are the more complex models we tested. 
\textbf{Second row:} Observed spectra ($S_1$ to $S_5$), shown as differences relative to $S_\mathrm{all}$ (colored points, binned in wavelength for visual clarity), and model fits ($\bar{M}_1$ to $\bar{M}_5$) relative to $\bar{M}_\mathrm{all}$ (solid colored lines). 
\textbf{Third row:} Same as the second row, but normalized at each wavelength to highlight relative changes.
\textbf{Bottom row:} Histograms of the posterior sample distributions for covering fractions in the 2-component fits.
        While the temperatures do not vary with time, the posterior estimates of the covering fractions evince a trend in which the covering fraction of the cool component ($f_1$) increases over the observations, reaching a maximum after the transits, and then decreases in the last time segment. 
        This is consistent with the stellar brightness variations.}
    }
    \label{fig:spectra_and_models}
\end{figure*}

\subsubsection{Stellar heterogeneity model and datasets}

We evaluated the evidence for stellar photospheric heterogeneity in the out-of-transit stellar spectra following the approach of \citet{Narrett2024} and utilizing the methods introduced by \citet{rackham:2024}. Similar to \citet{rackham:2024}, we considered composite stellar photospheres comprised of one to four spectral components. To generate the model stellar spectra for this study, we used the {\tt{SPHINX}} \citep{Iyer2023} spectral models. A linear interpolation between grid points was performed based on $T_\mathrm{eff}$, $\mathrm{[Fe/H]}$, and $\log g$, utilizing the \texttt{speclib} package\footnote{\url{https://github.com/brackham/speclib}} \citep{speclib-0.0-beta.0}. 
$\mathrm{[Fe/H]}$ and $\log g$ were fixed to literature values \citep{Agol2021}, leaving $T_\mathrm{eff}$ to parameterize the spectra.
As with previous studies \citep{lim:2023, Narrett2024, rackham:2024}, we include the 2MASS fluxes \citep{skrutskie:2006} as a constraint in the model while also fitting for a scale factor that applies to the JWST fluxes to account for any shortcomings in the calculation of the absolute flux level of the JWST spectra.

Given the extended out-of-transit baseline, we explored the time-dependence of stellar surface heterogeneities by repeating this analysis on mean stellar spectra derived from different subsets of the out-of-transit baseline. We split the pre-transit baseline into three equal-length subsets ($S_1$, $S_2$, $S_3$) and the post-transit baseline into two equal-length subsets ($S_4$, $S_5$; \autoref{fig:white_light}). We also analyzed the mean stellar spectrum derived from the full out-of-transit baseline ($S_\mathrm{all}$). Importantly, unlike \citet{rackham:2024}, we do not inflate the uncertainties on the JWST spectra; this leads to parameter inferences that do not account for the model-driven uncertainty (reflecting their lack of fidelity), but allows us to better investigate temporal trends in the inferences, the primary aim of the analysis here. The changes in the $S_1$ to $S_5$ spectra being well captured by the best-fit models support this approach (details in the next subsection).

\subsubsection{Photospheric inferences} \label{Photospheric inferences}

\autoref{fig:spectra_and_models} shows the results of the stellar spectral model fitting. We focus on the results of the 1-component and 2-component models hereafter, as they are the simplest models that capture the relevant behavior of the model fits. Comparing the $S_1$ to $S_5$ spectra (left column) shows that the star dimmed over the course of the observation, reaching a minimum just after the transit of b ($S_4$) before beginning to brighten again. This variability was most pronounced at ${\sim}1\,\micron$ in an absolute sense. In a relative sense, it was most pronounced at the shortest wavelengths probed (${\sim}0.6\,\micron$).

The best-fit models capture this behavior well. In the case of the 2-component model, the $S_\mathrm{all}$ spectrum is fit by 55.6\% coverage of a relatively cool $T_1 \approx 2000\,$K component\footnote{The lower temperature limit of the SPHINX model grid is 2000\,K, and so this represents a temperature upper limit for the cool component.} and 44.4\% coverage of a relatively warm $T_2 \approx 2600\,$K component (\autoref{fig:spectra_and_models}). Over the course of the observation, the coverage of the cool component increases from 55.5\% ($S_1$) to 55.7\% ($S_4$) before decreasing slightly again. Thus, our results suggest that the photosphere of the star is dominated by some relatively cool component, and its changing surface coverage can account for the variability of the out-of-transit spectra.

We note that our best-fit, two-component model yields an effective temperature (${\sim}2324$\,K) that is less than estimates of the stellar effective temperature in the literature \citep[$2566 \pm 26$\,K,][]{Agol2021}.
However, taking into account the systematic scale factor in our fit, the scaled flux of our model corresponds to an effective temperature of 2463\,K.
If we additionally take the systematic uncertainty on effective temperature to be 50\,K, given the limitations in stellar models of ultracool dwarfs \citep[e.g.,][]{Iyer2023, rackham:2024}, our model is consistent with previous estimates of the stellar effective temperature at the $2\sigma$ level.

\subsection{In-transit spectra}\label{sec:in-transit_spec}
To extract the transit spectra of both planets b and c, we first performed a joint fit to the white light curve. For both planets, we fitted the parameters time of transit ($t_0$), period ($P$), planet-to-star radius ratio ($R_p/R_\star$), semi-major axis in units of stellar radii ($a/R_\star$), and the inclination ($i$). Given the similar impact parameters of the two planets, we opted to fit for shared quadratic limb darkening coefficients\correction{, sampled using the parameterization described by \citet{Kipping2013}.} \correction{NIRSpec PRISM data often exhibit a linear decrease in flux over time, so we also included two parameters to describe this linear trend. Additionally, we included a Gaussian Process (GP) utilizing a squared exponential kernel to address residual correlated noise. This kernel is characterized by two parameters: the length scale, which determines the typical timescale over which data points are correlated (with larger length scales capturing smoother variations and smaller ones capturing rapid changes), and the amplitude, which sets the maximum covariance between data points. Additionally, we included a white noise parameter to account for additional uncorrelated noise and to prevent overfitting.} This resulted in a total of 17 free parameters, which were explored using an MCMC approach, implemented in Python through the \texttt{emcee} package \citep{Foreman-Mackey2013}. \correction{To make the GP calculation tractable, we binned the light curve using a bin size of 30 integrations, resulting in a time step of 40.8 seconds between bins and reducing the original 10,208 integrations to 341.}

\begin{figure*}[ht!]
	\centering
	\includegraphics[width=1.0\linewidth]{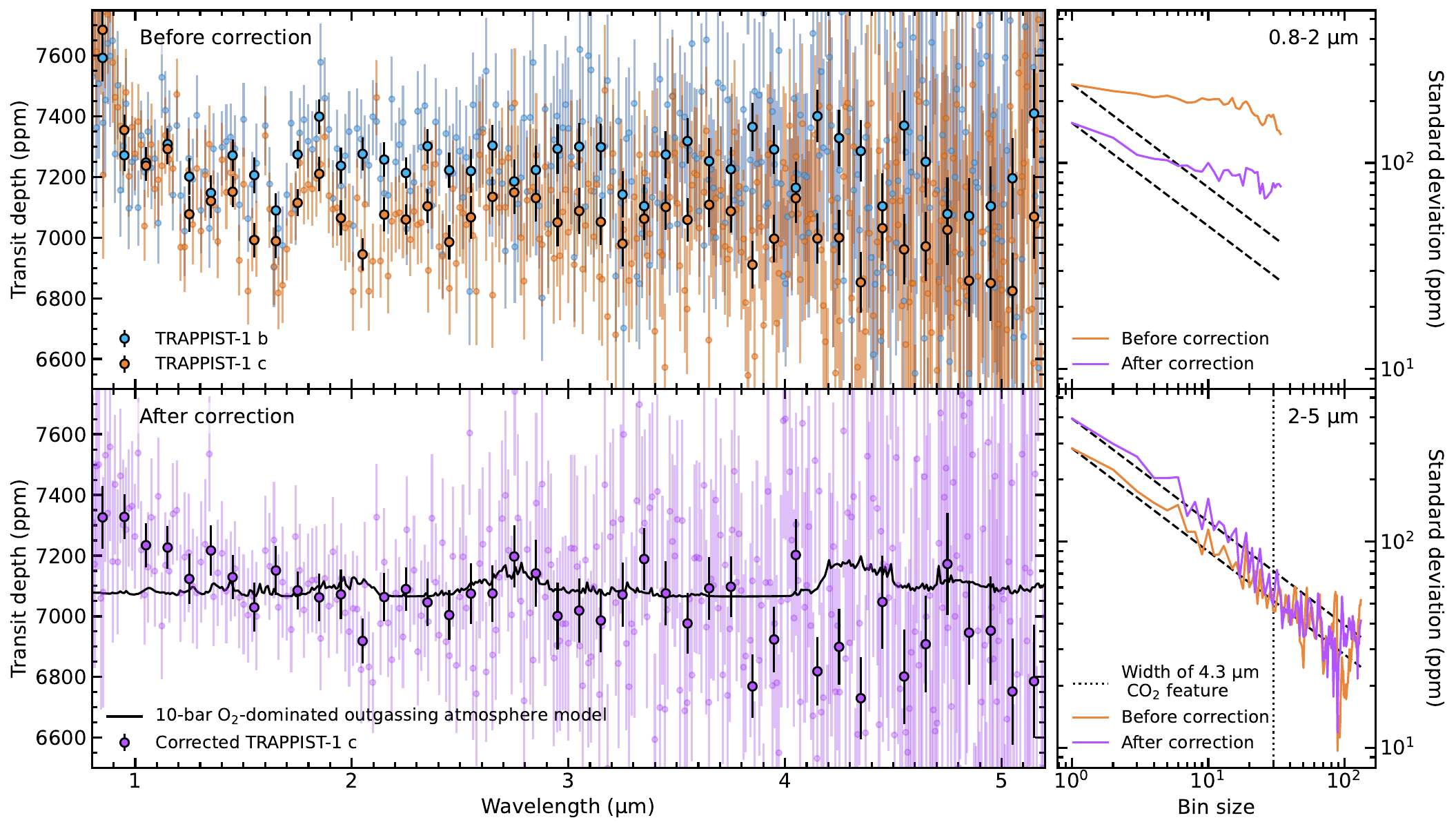}
	\caption{\correction{Transit spectra of TRAPPIST-1~b and TRAPPIST-1~c and the impact of the correction process. \textbf{Top left panel:}} Transit spectra for TRAPPIST-1~b (blue) and TRAPPIST-1~c (orange), shown both at native pixel resolution (faded points) and binned into 0.1~$\upmu$m intervals (solid points). \correction{\textbf{Bottom left panel:}} Corrected transit spectrum of TRAPPIST-1~c (purple). The corrected spectrum was computed as the ratio of the TRAPPIST-1~c transit spectrum to the TRAPPIST-1~b transit spectrum, multiplied by the weighted mean of the TRAPPIST-1~b transit depth in the 2--5~$\upmu$m range. The corrected spectrum is displayed at native pixel resolution and binned into 0.1~$\upmu$m intervals. Also plotted is a reference model for a 10-bar O$_2$-dominated outgassing atmosphere \citep[similar to that of][]{Lincowski2018}, illustrating potential atmospheric signatures and their expected amplitude in this wavelength range. \correction{\textbf{Top right panel:} Allan variance plot for the 0.8--2~$\upmu$m range, comparing the standard deviation of residuals before correction (orange) and after correction (purple) as a function of bin size for TRAPPIST-1~c. Dashed lines represent the expected noise reduction for uncorrelated (white) noise. \textbf{Bottom right panel:} Similar to the panel above it but for the 2--5~$\upmu$m range.}}
	\label{fig:transit_spectra}
\end{figure*}

We set Gaussian priors on $P$, $a/R_\star$, and $i$ for both planets, using the results from \cite{Agol2021}. We set uniform priors for the times of transit, planet radii, limb darkening parameters, and linear coefficients with wide but reasonable limits to allow for faster convergence of the MCMC chains. Finally, we set natural logarithm uniform priors for the GP hyperparameters. \correction{The prior on the GP length scale was chosen to be flexible enough to reflect the expected timescale of correlations in the systematics and ranged from $-6.05$ to $-2.52$, corresponding to $3.4~\text{minutes}$ to $1.9~\text{hours}$ (i.e., five times the time step between binned data points and half the duration of the full observation). The GP amplitude prior spanned $-27.63$ to $-13.82$, corresponding to a covariance range from $1~\text{ppm}$ to $1000~\text{ppm}$. The prior on the white noise parameter ranged from $-30$ to $-15$, corresponding to $0.3~\text{ppm}$ to $553~\text{ppm}$.} The full information on the priors and the resulting posteriors are displayed in \autoref{tab:fit_info}. 

Following the white-light curve fit, we proceeded to fit each individual dispersion bin, fixing the wavelength-independent parameters to those found in the white-light curve fit. We also fixed the hyperparameters of the GP to the values found in the white-light curve fit in order to prevent over-fitting the much lower SNR individual wavelength light curves. This leaves planet radii, the coefficients for the linear trend, and the two limb darkening parameters as free parameters in the fits to the light curves of the individual wavelength bins. The resulting transit spectra for both planets are shown in \autoref{fig:transit_spectra}.

A quantitative analysis of the transit spectra (see top left panel of \autoref{fig:transit_spectra}) reveals signs of stellar contamination for both TRAPPIST-1~b and TRAPPIST-1~c. This is evident from both the slope observed in the blue end of the spectrum \correction{(between 0.8 and 1.5 $\upmu$m)}, and from the water absorption clearly visible as two prominent features at 1.4 and 1.8\,$\upmu$m. Assuming that the stellar contamination affects the transit spectra of both planets identically, we can correct for this effect by taking the ratio of their spectra. We apply this correction to the transit spectrum of TRAPPIST-1~c by multiplying the weighted mean transit depth of TRAPPIST-1~b in the 2-5\,$\upmu$m range by the ratio of the two transit spectra. The resulting corrected spectrum of TRAPPIST-1~c is shown in the bottom left panel of \autoref{fig:transit_spectra}.

The corrected spectrum indicates that the water absorption features are effectively removed, and much of the slope at the bluest wavelengths is also mitigated. However, some residual slope remains, suggesting that the slope is steeper for TRAPPIST-1~c than for TRAPPIST-1~b. \correction{This difference in slope may be caused by the evolving stellar surface coverage inferred from the out-of-transit spectra (\autoref{fig:spectra_and_models}), which show measurable changes even over the short observation timescales. These dynamic changes in the coverage of cooler and warmer regions can produce slight differences in the contamination signals for the two planets, offering a plausible explanation for the observed slope variation. Alternatively, the difference in the steepness of the blue slope may result from the slightly different transit chords of the two planets across the stellar disk, although the two planets have very similar impact parameters \citep{Agol2021}. Redward of $2.0~\upmu$m, the two transit spectra are consistent with each other to within 110~ppm (\(1.2\sigma\)) in $0.1~\upmu$m bins.}

\correction{To further evaluate the wavelength-dependent impact of stellar contamination and the effectiveness of our correction method, we examine Allan variance plots, which illustrate how the standard deviation of residuals varies as a function of bin size (\(N\)). The top right panel of \autoref{fig:transit_spectra} focuses on the 0.8--2.0\,$\upmu$m range, while the bottom right panel highlights the 2.0--5.0\,$\upmu$m range.}

\correction{In the 0.8--2.0\,$\upmu$m range (top right), the standard deviation of the residuals consistently decreases across all bin sizes after correction compared to the pre-correction transit spectra. This behavior confirms that the correction effectively reduces structured noise caused by stellar contamination, despite introducing additional white noise from planet~b. Quantitatively, the noise floor in this wavelength range for a bin size of 10 decreases from $\sim$200 ppm pre-correction to $\sim$80 ppm post-correction (top right panel of Figure \ref{fig:transit_spectra}), representing a reduction in red noise by a factor of 2.5. Both the pre- and post-correction curves deviate from the theoretical white noise scaling (\(\propto 1/\sqrt{N}\)), remaining flatter and elevated above the expected line. This can be visually observed in the transit spectra pre- and post-correction, suggesting that residual systematic noise persists, albeit at a reduced level following correction.}

\correction{In the 2.0--5.0\,$\upmu$m range (bottom right), the variance of the unbinned residuals is higher post-correction, reflecting the addition of photon noise from planet~b, which is more significant in this wavelength range due to the lower SNR at longer wavelengths. Given that this study is based on a single transit observation, the spectrum at longer wavelengths is dominated by white noise. This dominance makes it challenging to robustly assess the degree of stellar contamination in the red portion of the spectrum.}

\correction{However, the out-of-transit analysis (see Section \ref{Photospheric inferences}) shows that stellar surface evolution is evident over the observation timescale, as shown in Figure \ref{fig:spectra_and_models}. While the relative changes during observations remain most pronounced at shorter wavelengths, variations are present across the entire probed wavelength range, including longer wavelengths, as illustrated in the third row of Figure \ref{fig:spectra_and_models}. Although the stellar variations seen in the out-of-transit spectra are not direct evidence of stellar contamination, they show a highly dynamic stellar surface throughout the spectrum, including at the redder wavelengths. Because transmission spectroscopy relies on relative changes at each wavelength, this implies that contamination also occurs at longer wavelengths.}

\correction{For the single transit presented in this work, stellar contamination is more difficult to discern at longer wavelengths in the transit spectra because of lower stellar flux and correspondingly higher noise. However, the same principles that enable effective correction in the blue regime should also apply in the red. At shorter wavelengths, where the noise is lower, the correction demonstrates its efficacy: the water absorption features disappear and the blue slope between 0.8--1.5\,$\upmu$m is significantly reduced. Assuming the correction is consistent across the spectrum, contamination at longer wavelengths should also be mitigated to a similar extent. Our best-fit stellar contamination model to the uncorrected transit spectra of TRAPPIST-1 c (see Figure \ref{fig:contamination_models}) suggests a contamination slope of approximately 300 ppm in the 0.8--1.5\,$\upmu$m range. At the same time, it also suggest a smooth contamination feature of \(\sim 200\) ppm between 4 and 5\,$\upmu$m. If the correction works uniformly across the spectrum, we expect the contamination amplitude in this red region to also be reduced by a factor of 2.5 (i.e., down to $\sim$ 80 ppm).}

\correction{The practical implications of this correction are significant. By shifting the noise source from a combination of structured red noise (stellar contamination) and white noise to one dominated by white noise, the challenge becomes one of achieving sufficient signal-to-noise through repeated observations rather than contending with systematic effects. While the white noise is inherently higher post-correction due to the photon noise contribution from planet~b, this trade-off is preferable for robust atmospheric characterization.}

\correction{It is worth noting that contamination features in the red are expected to be broader than the targeted CO$_2$ feature at 4.3\,$\upmu$m. Our best-fit contamination model predicts a relatively smooth signal of \(\sim 200\) ppm in this region expected down to $\sim$80~ppm after correction, which would not obscure sharp spectral features in the range of \(\sim 60\)–200 ppm expected for CO$_2$ absorption. Thus, while contamination remains a consideration, the correction technique demonstrated here enables us to focus on building sufficient signal-to-noise, allowing planetary absorption features to emerge prominently over the broad contamination signal.}

\correction{We note that the Allan variance plot for the 2--5 $\upmu$m range (bottom right panel of Figure~\ref{fig:transit_spectra}) presents a consistent picture. It shows that for a bin size of 30, which corresponds to the $\sim$0.3 $\upmu$m width of the most prominent CO\textsubscript{2} feature centered at 4.3 $\upmu$m, no structured noise above $\sim$80 ppm is found. In fact, it does not show a significant deviation from white noise behavior down to $\sim$ 40 ppm. This reinforces that, while white noise dominates one individual spectrum corrected for TLS at longer wavelengths, the absence of structured noise with large amplitudes ensures that planetary features such as the CO\textsubscript{2} signal should now be accessible by stacking TLS-corrected spectra via this method.}

\section{Discussion} \label{sec:discussion}

\correction{\subsection{A successful proof-of-concept \& implications for atmospheric searches}}

\correction{Our analysis demonstrates that back-to-back transits of TRAPPIST-1~b and c provide an effective method for mitigating stellar contamination. By leveraging these quasi-simultaneous transits, the correction reduces structured noise at shorter wavelengths ($<2~\upmu$m), successfully removing water absorption features and mitigating the blue slope, reducing its amplitude by approximately half. At longer wavelengths (2–5~$\upmu$m), contamination correction is more difficult to assess due to lower SNR, but the dynamic stellar surface variations observed in out-of-transit spectra suggest that contamination likely persists across the spectrum. This method thus transforms the dominant noise source from structured red noise to predominantly white noise, enabling planetary spectra to be reliably co-added over multiple epochs to improve precision.}

\correction{The observed similarity in stellar contamination signals for TRAPPIST-1~b and c, combined with their nearly identical radii, similar impact parameters \citep{Agol2021}, and position near the "dry side" of the cosmic shoreline \citep{Zahnle2017, Roettenbacher2017}, suggests that both planets are suitable reference points for stellar contamination correction in the TRAPPIST-1 system. Moreover, the interchangeability of TRAPPIST-1~b and c as reference planets increases observational flexibility by expanding the number of viable transit windows. For example, in JWST Cycle 3, c+e transit windows are twice as frequent as b+e windows with separations of less than 5 hours (see GO6456, PIs: Allen and Espinoza). However, while recent studies have ruled out a wide range of atmospheric scenarios for TRAPPIST-1~b and c, some atmospheric states remain viable given current data \citep{Greene2023, Zieba2023, Lincowski2023, Turbet2023, Ducrot2024, Radica2024}. Future studies should consider these possibilities when interpreting contamination corrections.}

\correction{The consistency in contamination signals observed in this study validates the use of back-to-back transits for epoch-specific contamination corrections and lays the groundwork for robust atmospheric characterization of the planets in the TRAPPIST-1 system. More broadly, this method provides a framework for addressing stellar contamination in other multi-planet systems with similar properties.}

\subsection{Insights from in- and out-of-transit spectra}

\begin{figure*}[t!]
    \centering
    \includegraphics[width=0.95\linewidth]{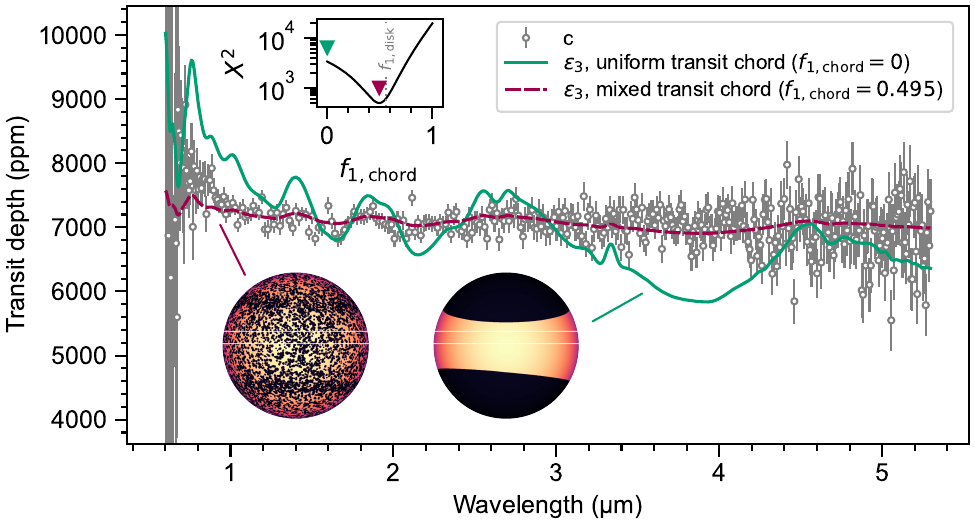}
    \caption{
        Comparison of uncorrected transit spectra and stellar contamination factors derived from out-of-transit stellar spectra.
        The transit spectrum of planet c is shown along with stellar contamination models calculated from the mean stellar spectrum just before the transit of c ($S_3$).
        The green line shows the model that assumes the entirety of the cooler spectral component is present outside of the transit chord.
        The purple line shows the best-fit stellar contamination model, allowing for a mixture of cool and warm spectral components throughout the stellar disk and a slight overdensity of the cool component outside the transit chord.
        Both contamination models have been multiplied by a flat planetary transit spectrum that yields the best fit to the data.
        The inset panel shows the $X^2$ value for models with varying levels of coverage of the cool component within the transit chord ($f_{1, \mathrm{chord}}$).
        Rather than the cool component being a monolithic unocculted feature, the relatively muted features in the transit spectrum point to the component spectra being relatively well-mixed across the stellar surface, with the cool component being slightly underrepresented in the transit chord.
        While this technique cannot probe the spatial distribution of features beyond their full-disk and transit-chord covering fractions, we present illustrative examples of stellar photospheres with spot distributions that reflect these models, with the transit chord of c highlighted \citep[produced with \texttt{spotter,}][]{spotter}.
    }
    \label{fig:contamination_models}
\end{figure*}

While we performed separate analyses of the in- and out-of-transit spectra, they contain related pieces of information. Indeed, the out-of-transit spectra depends on the level of photospheric heterogeneity, which also informs the stellar-contamination signal in transit. In this subsection, we combine the insights gained from each set of spectra in order to further constrain TRAPPIST-1's heterogeneities. 

To this end, we first calculated the stellar contamination signals ($\epsilon$) corresponding to the estimated level of photospheric heterogeneity and compared them to the uncorrected transit spectra. As an example, we focus on the transit spectrum of planet~c and the contamination signal ($\epsilon_3$) corresponding to the stellar spectrum observed just before its transit ($S_3$).

First, we calculated $\epsilon_3$ by assuming that the planet transits the warm component and the cool component is fully outside the transit chord (i.e., $f_{1, \mathrm{chord}} = 0$). As shown in \autoref{fig:contamination_models}, this results in a contamination model that increases at short wavelengths qualitatively in line with the transit spectrum but has notably stronger spectral features than the data.

For this reason, we also calculated a range of $\epsilon_3$ models that allowed the warm and cool components to be mixed throughout the photosphere and for their covering fractions within the transit chord to depart from their full-disk coverages.
Analytically, we modeled this as

\begin{equation}
    \epsilon = \frac{f_{1, \mathrm{chord}} F_1 + f_{2, \mathrm{chord}} F_2}
                    {f_{1, \mathrm{disk}} F_1 + f_{2, \mathrm{disk}} F_2}
\end{equation}
in which $F_1$ and $F_2$ are the spectra of the dominant ($\approx 2000$\,K) and non-dominant ($\approx 2600$\,K) components, respectively, $f_{1, \mathrm{disk}}$ and $f_{2, \mathrm{disk}}$ are their full-disk covering fractions, and $f_{1, \mathrm{chord}}$ and $f_{2, \mathrm{chord}}$ are their coverages within the transit chord.
We find that the model with $f_{1, \mathrm{chord}} = 0.495$ produces the best match to the data (\autoref{fig:contamination_models}), which reflects 6.1\% less coverage of the cool component within the transit chord than on the full disk.
In other words, reconciling our inference of the highly heterogeneous stellar photosphere from the stellar spectrum analysis with the strength of the contamination features evident in the transit spectrum requires that the two spectral components are generally well mixed over the photosphere of the star, though there is a relatively minor underabundance of the cool component within the transit chord.

In effect, this interpretation of a well-mixed heterogeneous photosphere for TRAPPIST-1 means that, when flares do not impact observations, the stellar contamination signal is notably smaller at long wavelengths than previous estimates that assumed an inhomogeneity fully outside the transit chord \citep[e.g.,][]{Rackham2018}.
Put quantitatively, at wavelengths ${>}2.5\,\micron$ the peak-to-peak amplitude of our model with a uniform transit chord and fully unocculted inhomogeneity is 1740\,ppm (green line in \autoref{fig:contamination_models}), while it is 280\,ppm for our best-fit model including the well-mixed photosphere (purple line).
This six-fold reduction in the expected scale of the stellar contamination signals at long wavelengths bodes well for the prospects of efforts to characterize any planetary atmospheres in the TRAPPIST-1 system at these wavelengths, which are further improved by the quasi-simultaneous transit technique we employ here.

\correction{Our work adds to the substantial and growing body of research that has attempted to characterize the photospheric heterogeneity of TRAPPIST-1. While various constraints have been offered based on inferences from the transit spectra of the TRAPPIST-1 planets alone \citep{Ducrot:2018, burdanov:2019, Agol2021, Radica2024}, multi-band photometry of the host star \citep{Morris2018}, or a combination of the transit spectra and normalized stellar spectra \citep{zhang:2018}, we should expect the clearest insights to come from analyses of the flux-calibrated spectrum of TRAPPIST-1 itself. Studies with such analyses have found a variety of inferences, such as a photosphere comprised of $\sim$62\%, $\sim$35\%, and $<$3\% coverages of $\sim$2400 K, $\sim$3000 K, and $\sim$5800 K components, respectively \citep{wakeford:2019}, or a photosphere comprised of $\sim$85\%, $\sim$15\%, and 185 ppm coverages of $\sim$2500 K, $\sim$2000 K, and $\sim$4900 K components \citep{garcia:2022}. Broadly, these results point to a highly heterogeneous photosphere with discrete contributions from unique components in the 2000–3000 K range and possibly a minor contribution from a much hotter temperature component, likely related to the star’s flaring activity \citep{Morris2018}. Nonetheless, these same studies have consistently emphasized the limitations of relying on current stellar models for cool stars to infer the photospheric heterogeneity of TRAPPIST-1 and derive corrections for transmission spectra \citep{wakeford:2019, garcia:2022, lim:2023, Davoudi2024}.}

\correction{Our results are no exception, and thus underscore the importance of empirical techniques for deriving the stellar photospheric heterogeneity \citep{Berardo2024} and/or correcting for the stellar contamination signal. We stress that our inferences of the stellar heterogeneity are not definitive assessments of the true properties of the photosphere of TRAPPIST-1, given the noted limitations in stellar spectral models to describe late-M dwarfs at the precision of JWST spectra, but reflect our current understanding based on the available data and models. Nonetheless, our work offers notable steps forward in this line of study by exploring the relationship between the photospheric inferences and the stellar brightness as a function of time and using the inferences from the out-of-transit stellar spectra in tandem with the stellar contamination signal to improve constraints on the degree of mixing of the photospheric components across the projected stellar disk.}

\section{Conclusions}\label{sec:conclusion}

In this study, we demonstrated a proof-of-concept for a model-independent method to mitigate stellar contamination in exoplanet transit spectra. This technique is feasible for use in multi-planet systems that host at least one planet without an atmosphere. The main idea is that stellar contamination should be nearly identical between two planets if their transits occur close together in time, enabling a correction for the contamination signal \citepalias{TJCI2024}. We showcased this technique using back-to-back transits of TRAPPIST-1~b and~c observed on July 9, 2024, specifically for this purpose, and found the following.

\begin{enumerate}
\item \textbf{Consistent Stellar Contamination:} The transit spectra of both planets display highly consistent stellar contamination signals.

\correction{\item \textbf{Effective Noise Reduction:} By using TRAPPIST-1~b to correct the transit spectrum of TRAPPIST-1~c, we demonstrate a significant reduction in stellar contamination at shorter wavelengths ($<$ 2 $\upmu$m), where the SNR is highest. Specifically, we achieve $2.5\times$ reduction of stellar contamination in this range. At longer wavelengths (2--5 $\upmu$m), the extent of stellar contamination is uncertain due to the lower SNR from a single transit. However, our out-of-transit analysis reveals stellar variations across the entire spectrum, suggesting that contamination is also present in this range. Although harder to assess at these wavelengths, the correction's demonstrated effectiveness at shorter wavelengths supports the expectation that it works similarly at longer wavelengths (down to at least 80ppm). This correction technique thus reduces the challenge to primarily white noise, which can be addressed with additional observations.}

\correction{\item \textbf{Interchangeable Reference Planets:} TRAPPIST-1~b and c, with their similar contamination signals observed in this study, can both serve as effective reference planets for correcting stellar contamination in the search for atmospheres within the TRAPPIST-1 system. While the empirical similarity of their transit spectra supports this approach, it is important to acknowledge that recent studies suggest that some atmospheric scenarios cannot be entirely ruled out for these planets given the available data.}

\item \textbf{Insights into Stellar Heterogeneities:} Our analysis provided new insights into TRAPPIST-1's stellar heterogeneities, revealing cold (${\sim}2000$\,K) and warm (${\sim}2600$\,K) regions with variable surface coverage.

\end{enumerate}

These results pave the way for more precise characterization of exoplanetary atmospheres in multi-planet systems, as traditional contamination correction methods have proven insufficient to achieve the necessary precision for probing the atmospheres of Earth-sized planets with the fidelity provided by current stellar models for late M dwarfs.\\\\

This material is based upon work supported by the National Aeronautics and Space Administration under Agreement No.\ 80NSSC21K0593 for the program ``Alien Earths''.
The results reported herein benefited from collaborations and/or information exchange within NASA’s Nexus for Exoplanet System Science (NExSS) research coordination network sponsored by NASA’s Science Mission Directorate.
HDL acknowledges support from the Carlsberg Foundation, grant CF22-1254.
A.B.-A.’s contribution to this research was in part carried out at the Jet Propulsion Laboratory, California Institute of Technology, under a contract with the National Aeronautics and Space Administration (80NM0018D0004).

\vspace{5mm}
\facilities{JWST(NIRSpec)}

\software{
    \texttt{Frida},
    \texttt{jwst} \citep{bushouse2024},
    \texttt{speclib} \citep{speclib-0.0-beta.0},
    \texttt{spotter} \citep{spotter}
}

\appendix

\clearpage

\begin{deluxetable*}{lll}
\tablecaption{Priors and Posteriors from Light Curve Fit \label{tab:fit_info}}
\tablehead{
\colhead{Parameter} & \colhead{Prior} & \colhead{Posterior}
}
\startdata
\hline
\multicolumn{1}{l}{\textbf{Planet b}} \\
\hline
Midtransit time (BJD\_TDB) & $\mathcal{U}(2460501.395278, 2460501.415278)$ & $2460501.405464^{+0.000010}_{-0.000011}$ \\
Orbital Period (days) & $\mathcal{N}(1.510826, 0.000006)$ & $1.5108261^{+0.0000058}_{-0.0000058}$ \\
Planet-star radius ratio ($R_p/R_\star$) & $\mathcal{U}(0.03, 0.13)$ & $0.08507^{+0.00016}_{-0.00017}$ \\
Semi-major axis ($a/R_\star$) & $\mathcal{N}(20.843, 0.094)$ & $20.802^{+0.052}_{-0.078}$ \\
Inclination ($^{\circ}$) & $\mathcal{N}(89.778, 0.165)$ & $89.80^{+0.14}_{-0.12}$ \\
\hline
\multicolumn{1}{l}{\textbf{Planet c}} \\
\hline
Midtransit time (BJD\_TDB) & $\mathcal{U}(2460501.376598, 2460501.396598)$ & $2460501.385348^{+0.000011}_{-0.000012}$ \\
Orbital Period (days) & $\mathcal{N}(2.421937, 0.000018)$ & $2.4219371^{+0.0000180}_{-0.0000179}$ \\
Planet-star radius ratio ($R_p/R_\star$) & $\mathcal{U}(0.03, 0.13)$ & $0.08453^{+0.00016}_{-0.00017}$ \\
Semi-major axis ($a/R_\star$) & $\mathcal{N}(28.549, 0.129)$ & $28.422^{+0.073}_{-0.111}$ \\
Inclination ($^{\circ}$) & $\mathcal{N}(89.778, 0.118)$ & $89.83761834^{+0.09}_{-0.08}$ \\
\hline
\multicolumn{1}{l}{\textbf{Shared Parameters}} \\
\hline
Limb darkening coefficient $u_1$* & $\mathcal{U}(0.0, 1.0)$ (sampled as $q_1$) & $0.341^{+0.019}_{-0.019}$ \\
Limb darkening coefficient $u_2$* & $\mathcal{U}(0.0, 1.0)$ (sampled as $q_2$) & $0.146^{+0.038}_{-0.037}$ \\
Linear trend intercept & $\mathcal{U}(-0.01, 0.01)$ & $0.000316^{+0.000419}_{-0.000333}$ \\
Linear trend slope & $\mathcal{U}(-0.01, 0.01)$ & $-0.0020^{+0.0027}_{-0.0027}$ \\
GP white noise & $\mathcal{U}(-30, -15)$ & $-19.10^{+0.12}_{-0.11}$ \\
GP amplitude & $\mathcal{U}(-27.63, -13.82)$ & $-15.65^{+1.11}_{-1.08}$ \\
GP length scale & $\mathcal{U}(-6.05, -2.52)$ & $-5.67^{+0.42}_{-0.28}$ \\
Eccentricity & 0, fixed & - \\
Argument of periapsis (deg) & 90, fixed & - \\
\enddata
\footnotesize{* Limb darkening coefficients $u_1$ and $u_2$ were derived from the posteriors of sampled parameters $q_1$ and $q_2$, using the parameterization from \citet{Kipping2013}. The priors listed correspond to $q_1$ and $q_2$, while the posteriors are shown for $u_1$ and $u_2$.}
\end{deluxetable*}

\bibliography{bib}{}
\bibliographystyle{aasjournal}
\end{document}